\documentclass[conference]{IEEEtran}

\ifCLASSINFOpdf
 
\else
  
\fi

\usepackage{amsmath,amssymb,amsthm}
\usepackage{color}
\usepackage[noadjust]{cite}
\usepackage{graphicx}
\usepackage{epsfig}
\usepackage{breqn}
\usepackage[caption=false,font=footnotesize]{subfig}
\usepackage{array}

\newtheorem{theorem}{Theorem}[section]

\allowdisplaybreaks
\newcommand\given[1][]{\:#1\vert\:}

\begin{document}
\bstctlcite{IEEEexample:BSTcontrol}

\title{Secure and reliable connectivity in heterogeneous wireless sensor networks}

\author{\IEEEauthorblockN{Rashad Eletreby and Osman Ya\u{g}an}
\IEEEauthorblockA{Department
of Electrical and Computer Engineering and CyLab, \\
Carnegie Mellon University, Pittsburgh,
PA, 15213 USA\\
reletreby@cmu.edu, oyagan@ece.cmu.edu}}

\maketitle

\begin{abstract}
We consider wireless sensor networks secured by the heterogeneous random key predistribution scheme under an on/off channel model. The heterogeneous random key predistribution scheme considers the case when the network includes sensor nodes with varying levels of resources, features, or connectivity requirements; e.g., regular nodes vs. cluster heads, but does not incorporate the fact that wireless channel are unreliable. To capture the unreliability of the wireless medium, we use an on/off channel model; wherein, each wireless channel is either on (with probability $\alpha$) or off (with probability $1-\alpha$) independently. We present conditions (in the form of zero-one laws) on how to scale the parameters of the network model so that with high probability the network is $k$-connected, i.e., the network remains connected even if {\em any} $k-1$ nodes fail or leave the network. We also present numerical results to support these conditions in the finite-node regime.
\end{abstract}


\IEEEpeerreviewmaketitle

\section{Introduction}
\subsection{Motivation and Background}
Wireless sensor networks (WSNs) comprise of wireless-capable sensor nodes that are typically deployed randomly in large scale to meet application-specific requirements. They facilitate a broad range of applications including military, health, and environmental monitoring, among others \cite{Akyildiz_2002}. Due to the nature of those applications, a battery-powered sensor node is typically required to operate for a long period of time and such a stringent requirement on battery life severely limits the communication and computation abilities of WSNs. For instance, traditional security schemes that require high overhead are not feasible for such resource-constrained networks. However, WSNs are usually deployed in hostile environments and left unattended, thus they should be equipped with security mechanisms to defend against attacks such as node capture, eavesdropping, etc. Random key predistribution schemes were proposed to tackle those limitations, and they are currently regarded as the most feasible solutions for securing WSNs; e.g., see \cite[Chapter~13]{Raghavendra_2004} and \cite{camtepe_2005}, and references therein.

Random key predistribution schemes were first introduced in the seminal work of Eschenauer and Gligor \cite{Gligor_2002}. Their scheme, that we refer to as the EG scheme, operates as follows: before deployment, each sensor node is assigned a {\em random} set of $K$ cryptographic keys, selected uniformly at random and without replacement from a key pool of size $P$. After deployment, two nodes can communicate {\em securely} over an existing channel {\em if} they share at least one key. The EG scheme led the way to several other variants, including the $q$-composite scheme, the random pairwise scheme \cite{Haowen_2003}, and many others.

Recently, Ya\u{g}an \cite{Yagan/Inhomogeneous} introduced a new variation of the EG scheme, referred to as the heterogeneous key predistribution scheme. The heterogeneous scheme generalizes the EG scheme by considering the case when the network includes sensor nodes with varying levels of resources, features, or connectivity requirements (e.g., regular nodes vs. cluster heads); a situation that is likely to hold in many real-world implementations of WSNs \cite{Yarvis_2005}. The scheme is described as follows. Given $r$ classes, each node is independently classified as a class-$i$ node with probability $\mu_i>0$ 
for each $i=1,\ldots, r$. Then, class-$i$ sensors are each assigned $K_i$ keys selected uniformly at random from a key pool of size $P$. Similar to the EG scheme, nodes that share key(s) can communicate securely over an available channel after the deployment; see Section \ref{sec:Model} for details. 

Since then, the work on the heterogeneous scheme has been extended in several directions by the authors; e.g., see \cite{Rashad/Inhomo,EletrebyISIT,EletrebyAllerton,EletrebyCDC}. In particular, the reliability of secure WSNs under the heterogeneous key predistribution scheme has been studied in \cite{Rashad/Inhomo}. There we obtained the probability of the WSN remaining securely connected even when each wireless link fails with probability $1-\alpha$ independently from others. This is equivalent to studying the secure connectivity of a WSN under an on/off channel model, wherein each wireless channel is {\em on} with probability $\alpha$ and {\em off} with probability $1-\alpha$ independently from other channels. Authors showed that network reliability exhibits a threshold phenomena and established critical conditions on the probability distribution $\pmb{\mu}=\{\mu_1,\mu_2,\ldots,\mu_r \}$,
and for the {\em scalings} (as a function of network size $n$) of the key ring sizes $\pmb{K}=\{K_1,K_2,\ldots,K_r\}$, the key pool size $P$, and the channel parameter $\alpha$ so that the resulting WSN is securely $1$-connected with high probability. Although these results form a crucial starting point towards the analysis of the heterogeneous key predistribution scheme, the connectivity results given in \cite{Rashad/Inhomo} do not guarantee that the network would remain connected when sensors fail due to battery depletion, get captured by an adversary, or when they are {\em mobile} (leading to a change in their channel probabilities). Therefore, sharper results that guarantee network connectivity in the aforementioned scenarios are needed.

\subsection{Contributions}
The objective of our paper is to address the limitations of the results in \cite{Rashad/Inhomo}. In particular, we consider the heterogeneous key predistribution scheme under an on/off communication model consisting of independent wireless channels each of which is either on (with probability $\alpha$), or off (with probability $1-\alpha$). We derive conditions on the network parameters so that the network is $k$-connected with high probability as the number of nodes gets large. The $k$-connectivity property guarantees network connectivity despite the failure of any $k-1$ nodes (or, links) \cite{PenroseBook}. We remark that our results are also of significant importance to {\em mobile} WSNs, since for a $k$-connected mobile WSN, any $(k-1)$ nodes are free to move anywhere while the rest of the network remains at least $1$-connected.

Our approach is based on modeling the WSN by an appropriate random graph and then establishing scaling conditions on the model parameters such that it is $k$-connectivity with high probability as the number of nodes $n$ gets large. We remark that the heterogeneous key predistribution scheme induces an inhomogeneous random key graph \cite{Yagan/Inhomogeneous}, denoted by $\mathbb{K}(n,\pmb{\mu},\pmb{K},P)$, while the on-off communication model induces a standard Erd\H{o}s-R\'enyi (ER) graph \cite{ER}, denoted by $\mathbb{G}(n,\alpha)$. The overall network can therefore be modeled by the intersection of an inhomogeneous random key graph with an ER graph, denoted $\mathbb{K} (n;\pmb{\mu},\pmb{K},P) \cap \mathbb{G}(n;\alpha)$. Put differently, the edges in $\mathbb{K} (n;\pmb{\mu},\pmb{K},P) \cap \mathbb{G}(n;\alpha)$ represent pairs of sensors that share a key and have an available wireless channel in between; i.e., those that can communicate securely in one hop. We present conditions on how to scale the parameters of the intersection graph $\mathbb{K} (n;\pmb{\mu},\pmb{K},P) \cap \mathbb{G}(n;\alpha)$ so that it is $k$-connected with high probability when the number of nodes $n$ gets large. 

\subsection{Notation and Conventions}
All limiting statements, including asymptotic equivalences are considered with the number of sensor nodes $n$ going to infinity. The random variables (rvs) under consideration are all defined on the same probability triple $(\Omega,\mathcal{F},\mathbb{P})$. Probabilistic statements are made with respect to this probability measure $\mathbb{P}$, and we denote the corresponding expectation by $\mathbb{E}$. 
We say that an event holds with high probability (whp) if it holds with probability $1$ as $n \rightarrow \infty$. 
In comparing
the asymptotic behaviors of the sequences $\{a_n\},\{b_n\}$,
we use
$a_n = o(b_n)$,  $a_n=w(b_n)$, $a_n = O(b_n)$, $a_n = \Omega(b_n)$, and
$a_n = \Theta(b_n)$, with their meaning in
the standard Landau notation.

\section{The Model}
\label{sec:Model}
We consider a network consisting of $n$ sensor nodes labeled as $v_1, v_2, \ldots,v_n$. Each sensor node is classified as one of the $r$ possible classes according to a probability distribution $\pmb{\mu}=\{\mu_1,\mu_2,\ldots,\mu_r\}$ with $\mu_i >0$
for each  $i=1,\ldots,r$ and $\sum_{i=1}^r \mu_i=1$. Before deployment, each class-$i$ node selects $K_i$ cryptographic keys uniformly at random from a key pool of size $P$. 
Clearly, the key ring $\Sigma_x$ of node $v_x$ is a $\mathcal{P}_{K_{t_x}}$-valued rv where $\mathcal{P}_{K_{t_x}}$ denotes the collection of all subsets of $\{1,\ldots,P\}$ with exactly $K_{t_x}$ elements and $t_x$ denotes the class of node $v_x$. The rvs $\Sigma_1, \Sigma_2, \ldots, \Sigma_n$ are then i.i.d. with
\begin{equation}
\mathbb{P}[\Sigma_x=S \mid t_x=i]= \binom P{K_i}^{-1}, \quad S \in \mathcal{P}_{K_i}.
\nonumber
\end{equation}
After deployment, two sensor nodes that share at lease one cryptographic key in common can communicate securely over an existing communication channel.

Throughout, we let $\pmb{K}=\{K_1,K_2,\ldots,K_r\}$, and assume without loss of generality that $K_1 \leq K_2 \leq \ldots \leq K_r$. 
Consider a random graph $\mathbb{K}$ induced on the vertex set $\mathcal{V}=\{v_1,\ldots,v_n\}$ such that two distinct nodes $v_x$ and $v_y$ are adjacent in $\mathbb{K}$, denoted by the event $K_{xy}$, if they have at least one cryptographic key in common, i.e.,
\begin{equation}
K_{xy} :=\left[\Sigma_x \cap \Sigma_y \neq \emptyset\right].
\label{adjacency_condition}
\end{equation}
The adjacency condition (\ref{adjacency_condition}) describes the inhomogeneous random key graph  $\mathbb{K}(n;\pmb{\mu},\pmb{K},P)$ that has been introduced in \cite{Yagan/Inhomogeneous}.
This model is also known in the literature as
the {\em general random intersection graph}; e.g., see \cite{Zhao_2014,Rybarczyk,Godehardt_2003}. 

The inhomogeneous random key graph models the {\em secure} connectivity of the underlying WSN.
In particular, the probability $p_{ij}$ that a class-$i$ node and a class-$j$ have a common key, and thus are adjacent in $\mathbb{K}(n;\pmb{\mu},\pmb{K},P)$, is given by
\begin{equation}
p_{ij}= \mathbb{P}[K_{xy}] = 1-{\binom {P-K_i}{K_j}}\Bigg/{\binom {P}{K_j}}
\label{eq:osy_edge_prob_type_ij}
\end{equation}
as long as $K_i + K_j \leq P$; otherwise if $K_i +K_j > P$, we clearly have $p_{ij}=1$.
We also define the \textit{mean} probability $\lambda_i$ of edge occurrence for a class-$i$ node in $\mathbb{K}(n;\pmb{\mu},\pmb{K},P)$. With arbitrary nodes $v_x$ and $v_y$, we have
\begin{align}
\lambda_i=\sum_{j=1}^r p_{ij} \mu_j,  \quad i=1,\ldots, r,
 \label{eq:osy_mean_edge_prob_in_RKG}
\end{align}
as we condition on the class of node $v_y$.

The preceding notion of secure connectivity between two nodes does not incorporate the fact that wireless channels are unreliable. In particular, two nodes sharing a cryptographic key are not necessarily able to communicate with one another because of the unreliability of the wireless channel. In this work, we consider the communication topology of the WSN as consisting of independent 
channels that are either {\em on} (with probability $\alpha$) or {\em off} (with probability $1-\alpha$). 
In particular, let $\{B_{ij}(\alpha), 1 \leq i < j \leq n\}$ denote i.i.d Bernoulli rvs, each with success probability $\alpha$. The communication channel between two distinct nodes $v_x$ and $v_y$ is on (respectively, off) if $B_{xy}(\alpha)=1$ (respectively if $B_{xy}(\alpha)=0$). 
Although the on-off channel model could be deemed too simple, it captures the unreliability of wireless links and enables a comprehensive analysis of the properties of interest of the resulting WSN, e.g., its connectivity. It was also shown that on/off channel model provides a good approximation of the more realistic disk model \cite{Gupta99} in many similar settings and for similar properties of interest; e.g., see \cite{Yagan/EG_intersecting_ER,YaganPairwise,yagan2011designing}.
The on/off channel model induces a standard Erd\H{o}s-R\'enyi (ER) graph $\mathbb{G}(n;\alpha)$ \cite{Bollobas}, defined on the vertices $\mathcal{V}=\{v_1,\ldots,v_n\}$ such that $v_x$ and $v_y$ are adjacent, denoted $C_{xy}$, if $B_{xy}(\alpha)=1$.

We model the overall topology of a WSN by the intersection of an inhomogeneous random key graph $\mathbb{K}(n;\pmb{\mu},\pmb{K},P)$ and an ER graph $\mathbb{G}(n;\alpha)$. Namely, nodes  $v_x$ and $v_y$ are adjacent in $\mathbb{K} (n;\pmb{\mu},\pmb{K},P) \cap \mathbb{G}(n;\alpha)$, denoted $E_{xy}$, if and only if they are adjacent in both $\mathbb{K}$ \textit{and} $\mathbb{G}$. In other words, edges in the intersection graph
$\mathbb{K} (n;\pmb{\mu},\pmb{K},P) \cap \mathbb{G}(n;\alpha)$ represent pairs of sensors that
can securely communicate as they have i) a communication link in between
that is {\em on}, and ii) a shared cryptographic key.
Therefore, studying the connectivity properties of $\mathbb{K} (n;\pmb{\mu},\pmb{K},P) \cap \mathbb{G}(n;\alpha)$ amounts to studying the secure connectivity of heterogeneous WSNs under the on/off channel model.

Throughout, we denote the intersection graph $\mathbb{K} (n;\pmb{\mu},\pmb{K},P) \cap \mathbb{G}(n;\alpha)$ by  $\mathbb{H}(n;\pmb{\mu},\pmb{K},P,\alpha)$. To simplify the notation, we let $\pmb{\theta}=(\pmb{K},P)$, and $\pmb{\Theta}=(\pmb{\theta},\alpha)$. The probability of edge existence between a class-$i$ node $v_x$ and a class-$j$ node $v_y$ in $\mathbb{H}(n;\pmb{\Theta})$ is given by
\begin{equation} \nonumber
\mathbb{P}[E_{xy} \given[\Big] t_x=i,t_y=j]=\mathbb{P}[K_{xy} \cap C_{xy} \given[\big] t_x=i,t_y=j]=\alpha p_{ij}
\end{equation}
by independence. Similar to (\ref{eq:osy_mean_edge_prob_in_RKG}), the mean edge probability for a class-$i$ node in $\mathbb{H}(n;\pmb{\mu},\pmb{\Theta})$ as $\Lambda_i$ is given by
\begin{align} 
\Lambda_i = \sum_{j=1}^r \mu_j \alpha p_{ij} = \alpha \lambda_i, \quad i=1,\ldots, r.
\label{eq:osy_mean_edge_prob_in_system}
\end{align}

From now on, we assume that the number of classes $r$ is fixed and does not scale with $n$, and so are the probabilities $\mu_1, \ldots,\mu_r$. All of the remaining parameters are assumed to be scaled with $n$.

\section{Main Result and Discussion}
\label{sec:results}

\subsection{Result}
We refer to a mapping $K_1,\ldots,K_r,P: \mathbb{N}_0 \rightarrow \mathbb{N}_0^{r+1}$ as a \textit{scaling} (for the inhomogeneous random key graph) as long as the conditions
\begin{equation}
2 \leq K_{1,n} \leq K_{2,n} \leq \ldots \leq K_{r,n} \leq P_n/2
\label{scaling_condition_K}
\end{equation}
are satisfied  for all $n=2,3,\ldots$. Similarly any mapping $\alpha: \mathbb{N}_0 \rightarrow (0,1)$ defines a scaling for the ER graphs. As a result, a mapping $\pmb{\Theta} : \mathbb{N}_0 \rightarrow \mathbb{N}_0^{r+1} \times (0,1)$ defines a scaling for the intersection graph $\mathbb{H}(n;\pmb{\mu},\pmb{\Theta}_n)$ given that condition (\ref{scaling_condition_K}) holds. We remark that under (\ref{scaling_condition_K}), the edge probabilities $p_{ij}$ will be given by
(\ref{eq:osy_edge_prob_type_ij}).

We now present a zero-one law for the $k$-connectivity of $\mathbb{H}(n;\pmb{\mu},\pmb{\Theta})$. 
\begin{theorem}
\label{theorem:kconnectivity}
{\sl
Consider a probability distribution $\pmb{\mu}=\{\mu_1,\ldots,\mu_r\}$ with $\mu_i >0$ for $i=1,\ldots,r$ and a scaling $\pmb{\Theta}: \mathbb{N}_0 \rightarrow \mathbb{N}_0^{r+1} \times (0,1)$. Let the sequence $\gamma: \mathbb{N}_0 \rightarrow \mathbb{R}$ be defined through
\begin{equation}
\Lambda_1(n)=\alpha_n \lambda_1(n) = \frac{\log n + (k-1)\log \log n+\gamma_n}{n}, \label{scaling_condition_KG}
\end{equation}
for each $n=1,2, \ldots$. 

(a) If $\lambda_1(n)=o(1)$, we have
\begin{equation} \nonumber
\lim_{n \to \infty} \mathbb{P} \left[\mathbb{H}(n;\pmb{\mu},\pmb{\Theta}_n) \text{ is }k\text{-connected}  \right]= 0   \quad \text{ if } \lim_{n \to \infty} \gamma_n=-\infty
\end{equation}

(b) If 
\begin{align}
P_n &= \Omega(n), \label{eq:conn_Pn} \\
\frac{K_{r,n}}{P_n}&=o(1), \label{eq:conn_KrPn} \\
\frac{K_{r,n}}{K_{1,n}} &=o(\log n), \label{eq:conn_Kr_K1} 
\end{align}
we have
\begin{equation} 
\lim_{n \to \infty} \mathbb{P} \left[\mathbb{H}(n;\pmb{\mu},\pmb{\Theta}_n) \text{ is }k\text{-connected}  \right]=1  \quad \text{ if } \lim_{n \to \infty} \gamma_n=\infty.
\label{eq:kconn_OneLaw_Statement}
\end{equation}
}
\end{theorem}

Put differently, Theorem~\ref{theorem:kconnectivity} states that $\mathbb{H}(n;\pmb{\mu},\pmb{\Theta}_n)$ is $k$-connected whp if the mean degree of class-$1$ nodes, i.e., $n \Lambda_1(n)$, is scaled as $\left(\log n+(k-1) \log \log n+\gamma_n\right)$ for some sequence $\gamma_n$ satisfying $\lim_{n \to \infty} \gamma_n=\infty$. On the other hand, if the sequence $\gamma_n$ satisfies $\lim_{n \to \infty} \gamma_n=-\infty$, then whp $\mathbb{H}(n;\pmb{\mu},\pmb{\Theta}_n)$ is {\em not} $k$-connected. This shows that the critical scaling for $\mathbb{H}(n;\pmb{\mu},\pmb{\Theta}_n)$ to be $k$-connected is given by $\Lambda_1(n)=\frac{\log n+(k-1)\log \log n}{n}$, with the sequence $\gamma_n:\mathbb{N}_0 \rightarrow \mathbb{R}$ measuring the deviation of $\Lambda_1(n)$ from the critical scaling.

Under an additional condition; namely, $\lambda_1(n)=o(1)$, the scaling condition (\ref{scaling_condition_KG}) can be given by \cite[Lemma 4.2]{Yagan/Inhomogeneous}
\begin{equation}
\lambda_1(n) \sim \frac{K_{1,n} K_{\textrm{avg},n}}{P_n}
\label{eq:lambda_1_asymp}
\end{equation}
where $K_{\textrm{avg},n} = \sum_{j=1}^{r}\mu_j K_{j,n}$  denotes the {\em mean} key ring size in the network. This shows that the minimum key ring size $K_{1,n}$ is of significant importance in controlling the connectivity and reliability of the WSN; as explained previously, it then also controls the number of {\em mobile} sensors that can be accommodated in the network. For example, 
 with the mean number $K_{\textrm{avg},n}$ of keys per sensor is fixed, we see that  
 reducing $K_{1,n}$ by half means that the smallest $\alpha_n$ (that gives the largest link failure probability $1-\alpha_n$) for which the network remains $k$-connected whp  is increased by two-fold for any given $k$;
e.g., see Figure \ref{fig:3} for a numerical example demonstrating this.

The proof of Theorem~\ref{theorem:kconnectivity} is lengthy and technically involved. Therefore, we omit the details in this conference version. All details can be found in \cite{EletrebyYaganICCLong}

\subsection{Comments on the additional technical conditions}
In establishing the zero-law of Theorem~\ref{theorem:kconnectivity}, it is required that $\lambda_1(n)=o(1)$. This condition is enforced mainly for technical reasons for the proof of the zero-law to work. A similar condition was also required in \cite[Thm 1]{Jun/K-Connectivity} for establishing the zero-law for $k$-connectivity in the {\em homogeneous} random key graph \cite{yagan2012zero} intersecting ER graph. As a result of (\ref{eq:lambda_1_asymp}), this condition is equivalent to
\begin{equation}
K_{1,n} K_{\textrm{avg},n} = o(P_n).
\label{eq:extra_cond_1_equiv}
\end{equation}
We remark that, in real-world WSN applications the
key pool size $P_n$ is in orders of magnitude larger than any key ring size in the network \cite{Gligor_2002,DiPietroTissec}. This is 
needed to ensure the resilience of the network against adversarial attacks. Concluding, (\ref{eq:extra_cond_1_equiv}) (and thus $\lambda_1(n)=o(1)$) is indeed likely to hold in most applications.

Next, we consider conditions (\ref{eq:conn_Pn}), (\ref{eq:conn_KrPn}), and (\ref{eq:conn_Kr_K1}) that are needed in establishing the one-law of Theorem~\ref{theorem:kconnectivity}. Conditions (\ref{eq:conn_Pn}) and (\ref{eq:conn_KrPn}) are also needed in real-world WSN implementations in order to ensure the {\em resilience} of the network against node capture attacks; e.g., see \cite{Gligor_2002,DiPietroTissec}. For example, assume that an adversary captures a number of sensors, compromising all the keys that belong to the captured nodes. If $P_n = O(K_{r,n})$ contrary to (\ref{eq:conn_KrPn}), then the adversary would be able to compromise a positive fraction of the key pool (i.e., $\Omega(P_n)$ keys) by capturing only a constant number of sensors that are of type $r$. Similarly, if $P_n = o(n)$, contrary to (\ref{eq:conn_Pn}), then again it would be possible for the adversary to compromise $\Omega(P_n)$ keys by capturing only $o(n)$ sensors (whose type does not matter in this case). In both cases, the WSN would fail to exhibit the {\em unassailability} property \cite{MeiPanconesiRadhakrishnan2008,YM_ToN} and would be deemed as vulnerable against adversarial attacks.
We remark that both (\ref{eq:conn_Pn}) and (\ref{eq:conn_KrPn}) were required in \cite{Jun/K-Connectivity,Yagan/Inhomogeneous} for obtaining the one-law for connectivity and $k$-connectivity, respectively, in similar settings to ours.

Finally, the condition (\ref{eq:conn_Kr_K1}) is enforced mainly for technical reasons and limits the flexibility of assigning very small key rings to a certain fraction of sensors when $k$-connectivity is considered. An equivalent condition was also needed in \cite{Yagan/Inhomogeneous} for establishing the one-law for connectivity in inhomogeneous random key graphs. We refer the reader to \cite[Section 3.2]{Yagan/Inhomogeneous} for an extended discussion on the feasibility of (\ref{eq:conn_Kr_K1}) for real-world WSN implementations, as well as possible ways to replace it with milder conditions. 

We conclude by providing a concrete example that demonstrates how all the conditions required by Theorem \ref{theorem:kconnectivity}
can be met in a real-world implementation. Consider any number $r$ of sensor types, and pick any probability distribution $\pmb{\mu}=\{\mu_1, \ldots, \mu_r\}$ with $\mu_i > 0$ for all $i=1,\ldots, r$. For any channel probability $\alpha_n = \Omega(\frac{\log n}{n})$, set $P_n = n \log n$ and use 
\[
K_{1,n} = \frac{(\log n)^{1/2+\varepsilon}}{\sqrt{\alpha_n}} \quad \textrm{and} \quad K_{r,n} = \frac{(1 + \varepsilon)(\log n)^{3/2-\varepsilon}}{\mu_r \sqrt{\alpha_n}}
\]
 with any $\varepsilon > 0$. Other key ring sizes $K_{1,n} \leq K_{2,n}, \ldots, K_{r-1,n} \leq K_{r,n}$ can be picked arbitrarily. In view of Theorem \ref{theorem:kconnectivity} and the fact \cite[Lemma 4.2]{Yagan/Inhomogeneous} that $\lambda_1(n) \sim \frac{K_{1,n} K_{\textrm{avg},n}}{P_n}$, the resulting network will be $k$-connected whp for any $k=1, 2, \ldots$. Of course, there are many other parameter scalings that one can choose.
 
\subsection{Comparison with related work}
Our paper completes the analysis that we started in \cite{EletrebyISIT} concerning the minimum node degree of the intersection model $\mathbb{H} (n;\pmb{\mu},\pmb{\Theta})$. There, we presented conditions on how to scale the parameters of the intersection model $\mathbb{H} (n;\pmb{\mu},\pmb{\Theta})$ so that the minimum node degree is no less than $k$, with high probability when the number of nodes $n$ gets large. It is clear that a graph can not be $k$-connected if its minimum node degree is less than $k$. Thus, we readily obtain the zero-law of Theorem~\ref{theorem:kconnectivity} by virtue of the zero-law of the minimum node degree being no less than $k$ \cite[Theorem~3.1]{EletrebyISIT}. Our paper completes the analysis of \cite{EletrebyISIT} by means of establishing the one-law of $k$-connectivity; thus, obtaining a fuller understanding of the properties of the intersection model $\mathbb{H} (n;\pmb{\mu},\pmb{\Theta})$. In particular, it was conjectured in \cite[Conjecture~3.2]{EletrebyISIT}, that under some additional conditions, the zero-one laws for $k$-connectivity would resemble those for the minimum node degree being no less than $k$. In our paper, we prove that this conjecture is correct and provide the extra conditions needed for it to hold.

Our results also extend the work by Zhao et al. \cite{Jun/K-Connectivity} on the homogeneous random key graph intersecting ER graph to the heterogeneous setting. There, a zero-one law for the property that the graph is $k$-connected was established for $\mathbb{H}(n,K,P,\alpha)$. Considering Theorem~\ref{theorem:kconnectivity} and setting $r=1$, i.e., when all nodes belong to the same class and thus receive the same number $K$ of keys, our results recover Theorem~2 of Zhao et al. (See \cite[Theorems~2]{Jun/K-Connectivity}). 

In \cite{Yagan/Inhomogeneous}, Ya\u{g}an established a zero-one law for $1$-connectivity for the inhomogeneous random key graph $\mathbb{K}(n,\pmb{\mu},\pmb{K},P)$ under {\em full} visibility; i.e., when all pairs of nodes have a reliable communication channel in between. Our paper  extends these results by considering more practical WSN scenarios where the unreliability of wireless communication channels are taken into account through the on/off channel model. Also, we investigate the $k$-connectivity of the network for any non-negative constant integer $k$; i.e., by setting $k=1$ and $\alpha_n=1$ for each $n=2, 3, \ldots$, we recover Theorem~2 in \cite{Yagan/Inhomogeneous}. 

Finally, our work improves upon the results by Zhao et al. \cite{Zhao_2014} who established zero-one laws for the $k$-connectivity of inhomogeneous random key graphs (therein, this model was referred to as the general random intersection graph). Indeed, by setting $\alpha_n=1$ for each $n=2, 3, \ldots$, our result recovers the results in \cite{Zhao_2014}. We also remark that the additional conditions required by main results of \cite{Zhao_2014} render them inapplicable in practical WSN implementations. This issue is explained in details in \cite[Section 3.3]{Yagan/Inhomogeneous}.

\section{Numerical Results}
\label{sec:numerical}
In this section, we present numerical results to support Theorem~\ref{theorem:kconnectivity} in the finite node regime. In all experiments, we fix the number of nodes at $n = 500$ and the size of the key pool at $P = 10,000$. To help better visualize the results, we use the curve fitting tool of MATLAB.

\begin{figure}[t]
\centerline{\includegraphics[scale=0.45]{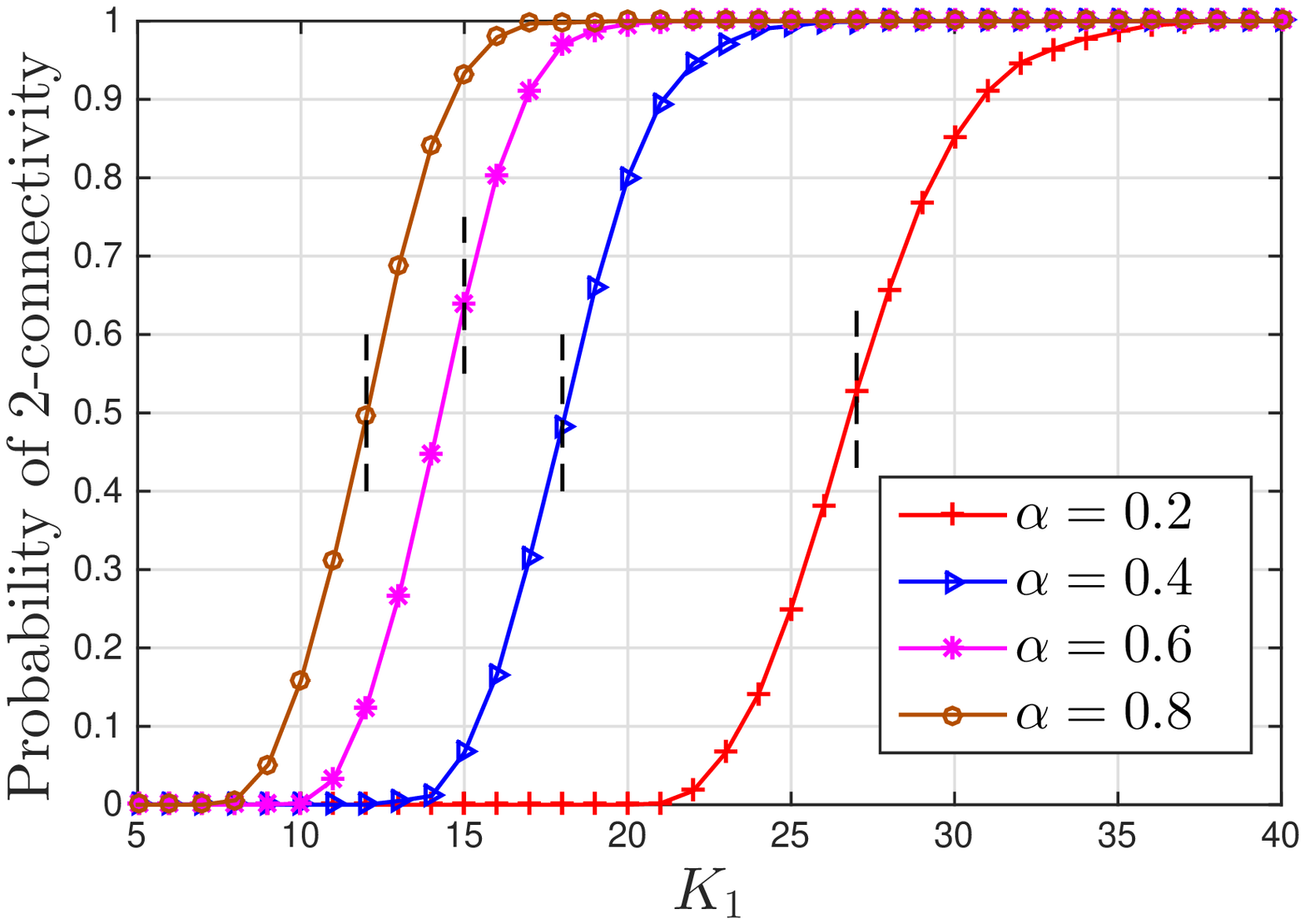}}
\caption{Empirical probability that $\mathbb{H}(n;\pmb{\mu},\pmb{\theta},\alpha)$ is $2$-connected as a function of $\pmb{K}$ for $\alpha = 0.2$, $\alpha = 0.4$, $\alpha = 0.6$, $\alpha = 0.8$ with $n = 500$ and $P = 10^4$; in each case, the empirical probability value is obtained by averaging over $200$ experiments. Vertical dashed lines stand for the critical threshold of connectivity asserted by Theorem~\ref{theorem:kconnectivity}.}
\label{fig:1}
\end{figure}

In Figure~\ref{fig:1}, we consider the channel parameters $\alpha = 0.2$, $\alpha = 0.4$, $\alpha = 0.6$, and $\alpha = 0.8$, while varying the parameter $K_1$, i.e., the smallest key ring size, from $5$ to $40$. The number of classes is fixed to $2$, with $\pmb{\mu}=\{0.5,0.5\}$. For each value of $K_1$, we set $K_2=K_1+10$. For each parameter pair $(\pmb{K}, \alpha)$, we generate $200$ independent samples of the graph $\mathbb{H}(n;\pmb{\mu},\pmb{\Theta})$ and count the number of times (out of a possible 200) that the obtained graphs are $2$-connected. Dividing the counts by $200$, we obtain the (empirical) probabilities for the event of interest.  

In Figure~\ref{fig:1} as well as the ones that follow we show the critical threshold of connectivity \lq\lq predicted" by Theorem~\ref{theorem:kconnectivity} by a vertical dashed line. More specifically, the vertical dashed lines stand for the minimum integer value of $K_1$ that satisfies
\begin{equation}
\hspace{-.6mm}\lambda_1(n)\hspace{-1mm}=\hspace{-1mm}\sum_{j=1}^2 \mu_j \hspace{-1mm}\left( \hspace{-.2mm} 1- \frac{\binom{P-K_j}{K_1}}{\binom{P}{K_1}}  \hspace{-.2mm}\right) \hspace{-.1mm}>  \hspace{-.1mm}\frac{1}{\alpha} \frac{\log n+(k-1)\log \log n}{n}
\label{eq:numerical_critical}
\end{equation}
with any given $k$ and $\alpha$.
We see from Figure~\ref{fig:1} that the probability of $k$-connectivity transitions from zero to one within relatively small variations in $K_1$. Moreover, the critical values of $K_1$ obtained by (\ref{eq:numerical_critical}) lie within the transition interval.  

\begin{figure}[t]
\centerline{\includegraphics[scale=0.45]{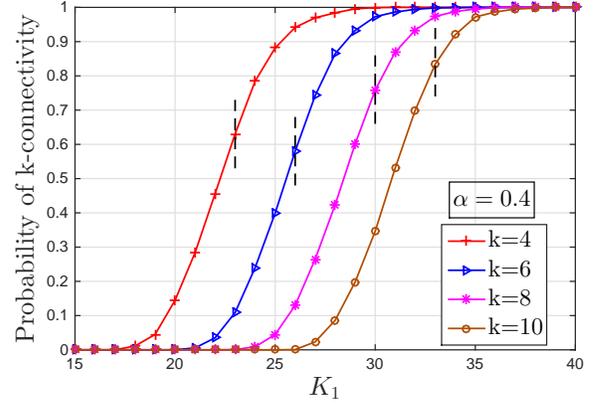}}
\caption{Empirical probability that $\mathbb{H}(n;\pmb{\mu},\pmb{\theta},\alpha)$ is $k$-connected as a function of $K_1$ for $k=4$, $k=6$, $k=8$, and $k=10$, with $n = 500$ and $P = 10^4$; in each case, the empirical probability value is obtained by averaging over $200$ experiments. Vertical dashed lines stand for the critical threshold of connectivity asserted by Theorem~\ref{theorem:kconnectivity}.}
\label{fig:2}
\end{figure}

In Figure~\ref{fig:2}, we consider four different values for $k$, namely we set $k = 4$, $k = 6$, $k = 8$, and $k=10$ while varying $K_1$ from $15$ to $40$ and fixing $\alpha$ to $0.4$. The number of classes is fixed to $2$ with $\pmb{\mu}=\{0.5,0.5\}$ and we set $K_2=K_1+10$ for each value of $K_1$. Using the same procedure that produced Figure~\ref{fig:1}, we obtain the empirical probability that $\mathbb{H}(n;\pmb{\mu},\pmb{\theta},\alpha)$ is $k$-connected versus $K_1$. The critical threshold of connectivity asserted by Theorem~\ref{theorem:kconnectivity} is shown by a vertical dashed line in each curve.
Again, we see that numerical results are in parallel with Theorem~\ref{theorem:kconnectivity}.

\begin{figure}[t]
\centerline{\includegraphics[scale=0.45]{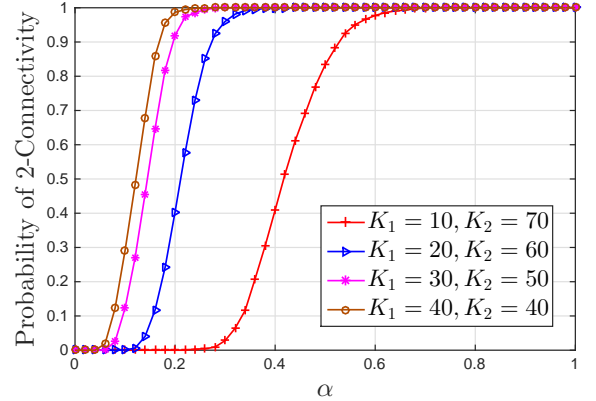}}
\caption{Empirical probability that $\mathbb{H}(n;\pmb{\mu},\pmb{\theta},\alpha)$ is $2$-connected with $n = 500, \pmb{\mu} = (1/2,1/2)$, and $P = 10^4$; we consider four choices of $\pmb{K} = (K_1,K_2)$ each with the same mean.}
\label{fig:3}
\end{figure}

Figure~\ref{fig:3} is generated in a similar manner with Figure~\ref{fig:1}, this time with an eye towards understanding the impact of the minimum key ring size $K_1$ on network connectivity. To that end, we fix the number of classes at $2$ with $\pmb{\mu}=\{0.5,0.5\}$ and consider 
four different key ring sizes $\pmb{K}$
each with mean $40$; we consider
$\pmb{K} = \{10,70\}$, $\pmb{K} = \{20,60\}$, $\pmb{K} = \{30,50\}$, and $\pmb{K} = \{40,40\}$.
We compare the probability of $2$-connectivity in the resulting networks while varying $\alpha$ from zero to one. We see that although the average number of keys per sensor is kept constant in all four cases, network connectivity improves dramatically as the minimum key ring size $K_1$ increases; e.g., with $\alpha=0.2$, the probability of connectivity is one when $K_1=K_2=40$ while it drops to zero if we set $K_1=10$ while increasing $K_2$ to $70$ so that the mean key ring size is still 40. 

\begin{figure}[t]
\centerline{\includegraphics[scale=0.45]{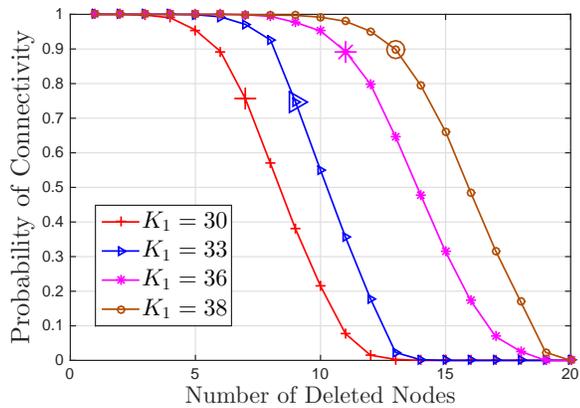}}
\caption{Empirical probability that $\mathbb{H}(n;\pmb{\mu},\pmb{\theta},\alpha)$ remains connected after deleting nodes from the {\em minimum vertex cut} set. We fix $n = 500, \pmb{\mu}=(1/2,1/2), \alpha=0.4,P = 10^4$, and choose $K_1$ and $K_2=K_1+10$
from (\ref{eq:numerical_critical}) for each  $k=8$, $k=10$, $k=12$, and $k=14$;
i.e., we use $K_1=30,33,36,38$, respectively.
}\label{fig:4}
\end{figure}

Finally, we examine the reliability of $\mathbb{H}(n;\pmb{\mu},\pmb{\theta},\alpha)$ 
by looking at the probability of 1-connectivity as the number of  deleted (i.e., failed) nodes increases. 
From a mobility perspective, this is equivalent to investigating the probability of a WSN remaining connected 
as the number of {\em mobile} sensors leaving the network increases. In Figure~\ref{fig:4}, we set $n = 500, \pmb{\mu}=\{1/2,1/2\}, \alpha=0.4,P = 10^4$, and select $K_1$ and $K_2=K_1+10$ from (\ref{eq:numerical_critical}) for
$k=8$, $k=10$, $k=12$, and $k=14$. With these settings, we would expect (for very large $n$) the network to remain connected whp after the deletion of up to 7, 9, 11, and 13 nodes, respectively.
Using the same procedure that produced Figure~\ref{fig:1}, we obtain the empirical probability that $\mathbb{H}(n;\pmb{\mu},\pmb{\theta},\alpha)$ is connected as a function of  the number of deleted nodes\footnote{We choose the nodes to be deleted from the {\em minimum vertex cut} of $\mathbb{H}$, defined as the minimum cardinality set whose removal renders it disconnected. This captures the worst-case nature of the $k$-connectivity property in a computationally efficient manner (as compared to searching over all $k$-sized subsets and deleting the one that gives maximum damage).} in each case.  We see that even with $n=500$ nodes, the resulting reliability is close to the levels expected to be attained asymptotically as $n$ goes to infinity. In particular, we see that the probability of remaining connected when $(k-1)$ nodes leave the network is around $0.75$ for the first two cases and around  $0.90$ for the other two cases.

\section*{Acknowledgment}
This work has been supported by the National Science Foundation through grant CCF-1617934,
and by the  Department of Electrical and Computer Engineering at Carnegie Mellon University.

\bibliographystyle{IEEEtran}
\bibliography{IEEEabrv,ICC}

\end{document}